\documentclass[prb,twocolumn,
showpacs,lengthcheck,superscriptaddress]{revtex4}
\usepackage{graphicx}
\usepackage{amsmath,amssymb}
\usepackage{bm}% bold math
\usepackage{ifthen}
\usepackage{dcolumn}% Align table columns on decimal point

\begin{document}

\title{Calculation of near-field 
scanning optical images of exciton, charged excition 
and multiexciton wavefunctions in
self-assembled InAs/GaAs quantum dots}
\author{Lixin He}
\affiliation{Laboratory of Quantum Information, University of Science 
and Technology of China, Hefei, Anhui 230026, Peoples Republic of China}
\author{Gabriel Bester}
\affiliation{National Renewable Energy Laboratory, Golden, Colorado 80401}
\author{Zhiqiang Su}
\affiliation{Laboratory of Quantum Information, University of Science 
and Technology of China, Hefei, Anhui 230026, Peoples Republic of China}
\author{Alex Zunger}
\affiliation{National Renewable Energy Laboratory, Golden, Colorado 80401}
\date{\today{}}% It is always \today, today,

\begin{abstract}

The near-field scanning optical 
microscopy images of excitonic wavefunctions in self-assembled
InAs/GaAs quantum dots are calculated using %single-particle 
an empirical pseudopotential method, followed by the
configuration interaction (CI) treatment of many-particle effects.
We show the wavefunctions of neutral exciton 
$X^0$ of different
polarizations, and compare them to those of the biexciton $XX$ and the charged
excitons $X^+$ and $X^-$.
We further 
show that the exciton $X(P_h \rightarrow S_e)$ transition which is 
forbidden in the far-field photoluminescence has comparable 
intensities to that of  
$X(S_h \rightarrow S_e)$ transition
in the near-field photoluminescence . 

\end{abstract}

\maketitle

\section{Introduction}

The 3D confinement and nearly perfect isolation from its environment
given in self-assemble semiconductor quantum dots (QDs)
leads to ``atom like'' electronic structure features manifested, among others, 
by $\mu$eV photoluminescence (PL) line widths
and a long coherence time. Such narrow and isolated levels
open the way for potential applications using as 
single-photon emitters \cite{michler00} 
and quantum-entangled sources. \cite{akopian06} At the same time, this allows
fundamental studies of $\mu$eV-scale fine-structure splitting
\cite{gammon96a,bayer99} of exciton lines
due to the electron-hole exchange interaction, spectral shift due to charged
excitons \cite{hartmann00,regelman01,urbaszek03}
as well as multi-exciton formation and decay.
However, the spatial resolution of the traditional 
far-field spectroscopy is limited to
$\lambda/2$, where $\lambda$ is the wavelength of the probing light, 
\cite{bryant98} 
much larger than the typical sizes of quantum dots
(a few tens to about a hundred nanometers).
Recently, it became possible to exam the
inner structures of an exciton in the QDs, and
map out its spatial distribution, 
using the near-field scanning optical 
microscopy (NSOM or SNOM). \cite{flack96,matsuda03}  
In the far field measurement, the transition intensity 
(within single-particle approximation) is proportional
to $|{\bf p}|^2
=|\langle \psi_c | {\bf r}| \psi_v \rangle |^2$, where $\psi_c$ and
$\psi_v$ are the conduction and valence band wavefunctions, respectively.
While in the far field measurements, the extent of the incident
electromagnetic wave  is such that its magnitude over the entire dot region
can be safely taken as constant, in the near-field, the electromagnetic
profile of the tip $ \xi({\bf r}- {\bf r}_0)$, has an extent comparable to the
size of the nanostructure and must be explicitly taken into account:
\begin{equation}
E({\bf r})= \xi({\bf r}- {\bf r}_0)E_0\, , 
\end{equation}
In this case, the transition intensity is modulated by 
the shape of the external field, i.e., 
proportional to $|{\bf p} ({\bf r_0})|^2
=|\langle \psi_c | {\bf r} \xi({\bf r}- {\bf r}_0) | \psi_v \rangle |^2$.
The resolution of NSOM can thus be
much higher than \cite{flack96} $\lambda/2$, 
providing details on how the
carriers are distributed in the quantum dots.

In a QD, the electron and hole wavefunctions can be approximated by 
products of envelope functions and Bloch functions. 
In the far field measurement, the interband transitions obey two kinds of 
selection rules:
one related to the Bloch part of the wavefunctions, and one to
the envelope component of the wavefunctions. 
For example, from the four possible $S_h \rightarrow S_e$ excitonic
transitions,
two are bright and two are dark. The dark transitions are forbidden because the
total angular momentum 
from the electron and hole {\it Bloch functions} parts sum up to
two, and can hence not be carried out by a single photon . 
In contrast, all of the eight $P_h \rightarrow S_e$ transitions are forbidden
because a photon does not couple effectively to the {\it envelope}
orbital angular momentum of the exciton. 
However, in NSOM, where the electromagnetic field is applied inhomogeneously
only to a part of the dot, 
the global spatial symmetry is broken, 
thus altering the selection rules for the envelope functions
and allowing transitions that are forbidden in far field measurement.
\cite{hohenester05}
We show here that the $P_h \rightarrow S_e$ transitions (which are far-field
forbidden) have comparable intensities in NSOM 
to the $S_h \rightarrow S_e$ transitions. 
Furthermore, NSOM provides additional far-field allowed
information about the Bloch part of the
wavefunctions, while the far field spectroscopy is lacking. 
For example, by comparing the transition 
intensities of different polarizations, one could determine the 
relative phase between 
$|X\rangle$ and $|Y\rangle$ components of the hole wavefunctions
at each point.  In contrast, in the far
field measurements, only the total polarization 
is measured and the wavefunction phase information of the Bloch
part is lost.

Recently, Matuda et al. \cite{matsuda03}
have measured the near-field PL image of naturally 
occurring GaAs QDs formed spontaneously in narrow quantum wells. 
The enhanced NSOM spatial resolution is as high as 30 nm, much smaller than
the measured QD sizes of $\sim$ 200 nm. 
Unfortunately, this spatial resolution is still too low for 
self-assembled InAs/GaAs quantum dots.
In this paper, we predict via atomistic pseudopotential method, 
the near-field PL images 
of the neutral exciton $X^0$, the biexciton $XX$ and the charged excitons 
$X^+$, $X^-$ in the much smaller 
self-assembled InAs/GaAs quantum dots, in anticipating of future NSOM
measurements.

\section{Methods}

\subsection{Far-field {\it vs} NSOM transition elements}

To obtain the NSOM images, one must first calculate
calculate the single-particle
wavefunctions for electrons and holes by solving the
Schr\"{o}dinger equations,
\begin{equation}
\left[ -{1 \over 2} \nabla^2 
+ V_{eff}({\bf r}) \right] \psi_i({\bf r})
=\epsilon_i \;\psi_i({\bf r}) \; ,
\label{eq:schrodinger}
\end{equation}
where $V_{eff}$ is the effective potential for the electrons in the
quantum dots.  For example, in an effective mass approximation (EMA), 
the effective potential is usually taken as
two-dimensional harmonic potential, whereas in atomistic approaches, 
\cite{williamson00} $V_{eff}$ is a superposition of screened atomic 
pseudopotentials.  
Once one has the electron and hole single-particle wavefunctions, it is
possible to
construct the {\it many-particle} wavefunctions of the excitonic states, e.g., 
via the configuration interaction (CI) method,
\cite{franceschetti99} by expanding them as the linear
combination of Slater determinants constructed from the single-particle 
electron and hole wavefunctions.
The ground state can be described by a single determinant,
\begin{eqnarray}
&&\Phi_0({\bf r}_1,\sigma_1, \cdots, {\bf r}_N,\sigma_N)\\ \nonumber
 &&=\mathcal{A} [\psi_1({\bf r}_1,\sigma_1)
\cdots\psi_v({\bf r}_v,\sigma_v)\cdots\psi_N({\bf r}_N,\sigma_N)]\, ,
\end{eqnarray}
where $N$ is the total number of electrons in the system,
$\sigma$ is the spin index, and $\mathcal{A}$ is the
anti-symmetrizing operator.
The Slater determinants of neutral excitons $\Phi_{v,c}$ are 
obtained by promoting an electron at the valence state $\psi_v$ to
the conduction state $\psi_c$ from the 
ground state $\Phi_0$, i.e.,
\begin{eqnarray}
&&\Phi_{v,c}({\bf r}_1,\sigma_1, \cdots, {\bf r}_N,\sigma_N)\\ \nonumber
&&=\mathcal{A} [\psi_1({\bf r}_1,\sigma_1)
\cdots\psi_c({\bf r}_c,\sigma_c)\cdots\psi_N({\bf r}_N,\sigma_N)]\, .
\end{eqnarray}
The $\alpha$-th neutral exciton wavefunction can be written as,
\begin{equation}
\Psi^{(\alpha)}_X= \sum_{v=1}^{N_v}\sum_{c=1}^{N_c}  
C_{v,c}^{(\alpha)}\, \Phi_{v,c}\, ,
\end{equation}
where $N_v$ and $N_c$ are the number of valence and conduction states
included in the expansion of the exciton wavefunctions respectively. 
The coefficients $C_{v,c}^{(\alpha)}$ are obtained by diagonizing the
many-particle Hamiltonian $H$ in the bassis set $\{ \Phi_{v,c} \}$. 
\cite{franceschetti99}

The far-field optical absorption of the $\alpha$ transition 
can be written as,
\begin{equation}
\sigma(\omega) \propto \sum_{\alpha} |{\bf M}^{(\alpha)}|^2
\delta(\hbar \omega -E^{(\alpha)} )\, ,
\label{eq:optic}
\end{equation}
where, ${\bf M}^{(\alpha)}$ are the dipole matrix
elements between initial many-particle
state $| \Psi_i \rangle$ and the final state  
$| \Psi_f \rangle$,
\begin{equation}
{\bf M}^{(\alpha)}=\langle \Psi_f\, |{\bf r} | \Psi_i \rangle \, .
\label{eq:M}
\end{equation}
$\omega$ and $E^{(\alpha)}$ are the frequency of the external 
E-field and the exciton transition energy, respectively.
For a neutral exciton, the dipole matrix
elements ${\bf M}$ can be written in terms of single-particle orbitals as,
\begin{equation}
{\bf M}^{(\alpha)}= \langle \Phi_0\, |{\bf r} | \Psi_X \rangle =
\sum_{v=1}^{N_v}\sum_{c=1}^{N_c}  C_{v,c}^{(\alpha)} \; 
\langle \psi_v |{\bf r} | \psi_c \rangle \; ,
\label{eq:M-X}
\end{equation}
where, ${\bf p}=\langle \psi_v |{\bf r} | \psi_c \rangle$ are 
the single-particle
dipole matrix elements between the conduction state $\psi_c$ and valence
state $\psi_v$.
Similarly, one can calculate ${\bf M}$ for charged excitons $X^+$, $X^-$ and
biexciton $XX$.

For NSOM, the external E-field is applied locally at 
${\bf r}_0$ modulated by the electromagnetic profile of the tip
$\xi({\bf r}- {\bf r}_0)$,
which can be taken as a Gaussian function, e.g.,
\begin{equation}
\xi({\bf r}- {\bf r}_0)
=c \;e^{-|{\bf r}-{\bf r}_0|^2/a^2}\; ,
\end{equation}
where $a$ is the width of the excitation field, much smaller than
the dimension of the QD.
For such local optical transition, we have to replace 
${\bf  M}^{(\alpha)}$ in Eq.~(\ref{eq:optic}) by
\begin{equation}
{\bf M}^{(\alpha)} ({\bf r}_0)= \sum_{v=1}^{N_v} \sum_{c=1}^{N_c} 
C_{v,c}^{(\alpha)} \; 
\langle \psi_v |{\bf r}\xi({\bf r}- {\bf r}_0)  | \psi_c \rangle \; ,
\label{eq:M-loc}
\end{equation}
equivalent to replacing in Eq. (\ref{eq:M-X}) 
the single-particle dipole moment 
${\bf p}=\langle \psi_v |{\bf r} | \psi_c \rangle$
by the local single-particle dipole moment 
${\bf p} ({\bf r}_0)=\langle \psi_v |{\bf r} \xi({\bf r}- {\bf r}_0) | \psi_c
\rangle$.

\begin{figure}
\includegraphics[width=3.0in,angle=0]{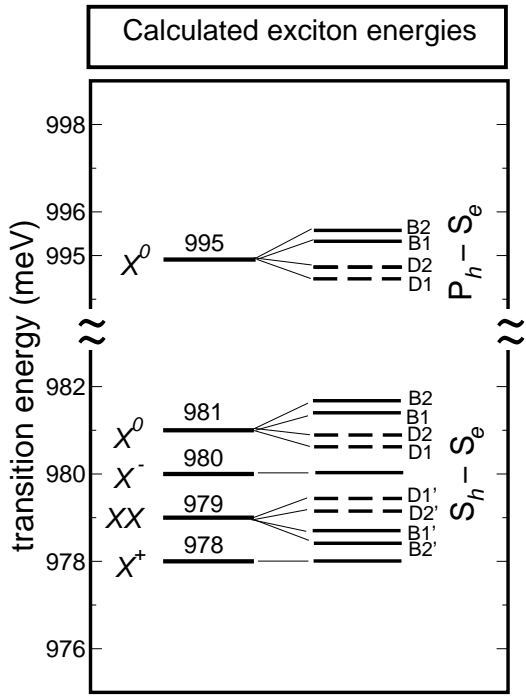}
\caption{The exciton energies of the lowest
neutral exciton $X^0$, the charged excitons $X^+$, $X^-$ 
and the biexciton $XX$ of self-assembled InAs/GaAs QD with 
base $b$=27.5 nm and height $h$=3.5 nm. 
The righthand side shows 
the schematic energy levels of fine structure splittings of each
state. The solid lines ($B1$, $B2$, etc.) denote bright ($B$) states,
whereas the dashed lines($D1$, $D2$, etc.) denote dark ($D$) states.}
\label{fig:energy}
\end{figure}

\subsection{Evaluation of the NSOM transition elements: 
an atomistic approach}

In the present work, we calculate the single-particle energy 
levels and wavefunctions 
in a atomistic screened pseudopotential scheme,
where the total potential $V_{eff}({\bf r})$ 
is a superposition of
local, screened atomic pseudopotentials $v_{\alpha}({\bf r})$, plus a nonlocal
spin-orbit potential $V_{\rm so}$ i.e.,
$V_{eff}({\bf r}) =\sum_{n,\alpha} 
v_{\alpha}({\bf r} - {\bf R}_{n,\alpha})+V_{\rm so}$.
The atomic positions $\{{\bf R}_{n,\alpha}\}$ are obtained from
minimizing the total bond-bending and
bond-stretching elastic energy 
using the Valence Force Field (VFF)
model.~\cite{keating66,martins84}
The atomistic pseudopotentials $v_{\alpha}$ ($\alpha$=In, Ga, As)
are fitted to the physically important quantities of {\it bulk} InAs and GaAs, 
including band energies, band-offsets,
effective masses, deformation potentials and alloy
bowing parameters, etc.~\cite{williamson00}
Equation (\ref{eq:schrodinger}) is solved  by expanding $\psi({\bf r})$
as the ``Linear Combination of Bulk Bands'' (LCBB)
of band index $m$
and wave vector ${\bf k}$ of material $\lambda$ (= InAs, GaAs), strained
uniformly to strain $\tensor{\epsilon}$ (following Ref.~\onlinecite{wang99b})
i.e.,
\begin{equation}
\psi({\bf r}) = \sum_{\bf k}\sum_{m}\, 
c^{(\lambda)}_{m, {\bf k}} u^{(\lambda)}_{m,{\bf k},
\tensor{\epsilon}}({\bf r}) 
e^{i{\bf k}\cdot{\bf r}} \, .
\label{eq:lcbbwavf}
\end{equation} 
We use the lowest conduction bands for electrons, 
and three highest valence bands for holes.
Instead of calculating the local dipole matrix elements using 
Eq. (\ref{eq:lcbbwavf}), 
we project $\psi({\bf r})$ onto (strained) Bloch functions of InAs at
the $\Gamma$ point in the Brillouin zone,
\begin{equation}
\psi({\bf r}) = \sum_{m=1}^{N_{\rm p}} f_{m}({\bf r}) u_{\Gamma,m} ({\bf r})
\, .
\label{eq:gamma-wavf}
\end{equation}
To simplify the notation, we drop the material index $\lambda$ and strain index
$\tensor{\epsilon}$ in Eq. (\ref{eq:gamma-wavf}).
Here, $N_{\rm p}$ is the number of the projection bands, which is usually much
larger than the number of bands used in the LCBB equation.
We found that $N_{\rm p}$= 9 provide accurate projections. 
Here, $f_{m}({\bf x})$ 
is the envelope function for the $m$-th band.
The envelope functions are slow varying functions, 
compared to the Bloch functions. We can therefore
separate the Bloch functions and envelope functions by dividing 
the volume into small regions, e.g. the 8-atom unit cells. 
In each unit cell, the envelope functions $f_{m}({\bf x})$ 
and $\xi({\bf r})$ can be treated as constants, i.e.,
\begin{eqnarray}
{\bf p} ({\bf r}_0) &=& \int d{\bf r} \psi^*_v ({\bf r})  \psi_c ({\bf r}) 
{\bf r} \xi({\bf r}- {\bf r}_0) \\ \nonumber
 & \approx & \sum_{{\bf r}_i}  \left[ \sum_{m,m'}
f^{*(v)}_{m} ({\bf r}_i)  f^{(c)}_{m'} ({\bf r}_i) 
 {\bf d}_{m,m'} \right]\;  \xi({\bf r}_i- {\bf r}_0)
\label{eq:dipole-simple}
\end{eqnarray}
where, ${\bf r}_i$ is the position of $i$-th cell and
\begin{equation}
{\bf d}_{m,m'}={1\over V} \int d{\bf r}\; u^*_{\Gamma,m} ({\bf r})\, {\bf r}\,
 u_{\Gamma,m'} ({\bf r}) \, ,
\label{eq:Mvicj}
\end{equation}
is the dipole matrix element between bulk bands.
In a periodic system, it is more convenient to use an alternative formula to
calculate the dipole, \cite{Dragoman_book}
\begin{equation}
{\bf d}_{m,m'} 
={1 \over im\omega_{m,m'}}
\langle u_{\Gamma,m} | {\bf \hat{p}} | u_{\Gamma,m'}\rangle\, . 
\label{eq:dmm} 
\end{equation}
If we assume high spatial NSOM resolution, i.e., 
$\xi({\bf r}_i- {\bf r}_0) =\delta({\bf r}_i- {\bf r}_0)$, 
Eq.~(\ref{eq:dipole-simple}) can be further simplified as,
\begin{equation}
{\bf p} ({\bf r}_0)
\approx  \sum_{m,m'}
f^{*(v)}_{m} ({\bf r}_0)  f^{(c)}_{m'} ({\bf r}_0) 
 {\bf d}_{m,m'} \, . 
\label{eq:dipole-simple-delta}
\end{equation}
Combining Eq. (\ref{eq:M-loc}) and 
Eq. (\ref{eq:dipole-simple-delta}), 
we see that
the shape of NSOM transition is the sum of the (multi-band) 
envelope functions of the exciton
weighted by the (single-particle) dipole matrix elements, i.e.,
\begin{equation}
{\bf M}^{(\alpha)} ({\bf r}_0)= 
\sum_{v=1}^{N_v} \sum_{c=1}^{N_c} 
C_{v,c}^{(\alpha)} \; \left[ \sum_{m,m'}  f^{*(v)}_{m} ({\bf r}_0)  
f^{(c)}_{m'} ({\bf r}_0) {\bf d}_{m,m'} \right] \, .
\label{eq:nsom-final}
\end{equation}

Previously, the NSOM images have been calculated 
via single-band effective mass approximation (EMA). 
\cite{nair97,hohenester05,hohenester04} 
In this approximation, the
dipole moment ${\bf d}$ is directly taken from bulk materials or 
from quantum wells, \cite{hohenester04} 
while in our case, it is calculated for nine bands in
Eq. (\ref{eq:dmm}) and 
is specific to the nanostructure considered.
In the single-band EMA,
only the envelope part of the information is kept in the
calculation of NSOM images,
whereas detailed information associated with the Bloch functions, 
such as polarizations is lost. 
In the more advanced k$\cdot$p methods, one includes
the heavy-hole, light-hole, and spin-orbit bands at $\Gamma$ 
point of the Brillouin zone in the calculations. 
For the direct gap
materials, these approximations give qualitative right results, 
but for indirect gap materials, such
Si quantum dots, one need to include many more bands.

\begin{figure}
\includegraphics[width=3.0in,angle=0]{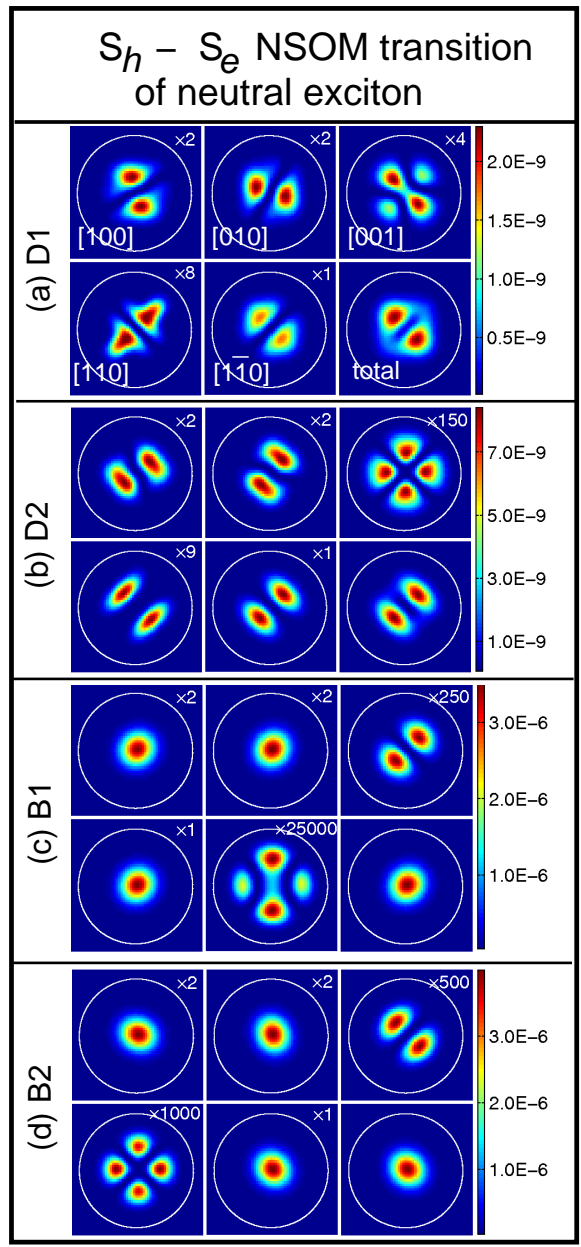}
\caption{ (Color online)
The NSOM of excitonic $S_h\rightarrow S_e$ transitions in 
self-assembled InAs/GaAs QD with $b$=27.5 nm and $h$=3.5 nm, 
calculated from Eq. (\ref{eq:nsom-final}).
The two dark excitons $D1$, $D2$ are plotted in (a) and (b) respectively,
whereas the two bright excitons $B1$, $B2$ are plotted in (c) and (d). 
The unit of the color-bar is arbitrary but equal for all sub-figures.
In each sub-figure, there are six panels showing
the transition intensities of different polarizations. 
In the first row, the
polarizations are along the [100], [010] and [001] direction respectively,
and in the
second row, the polarizations are along the [110] and [1$\bar{1}$0]
directions. The last panels show 
the sum of transition intensities of different polarizations.
We magnify the transition intensities 
by the factors on the upper-right corner of each
panel. 
The white circles (27.5 nm in diameter) show the boundaries the QDs.
} 
\label{fig:1e1h-ss}
\end{figure}

\section{Results and Discussion}

\subsection{The excitonic spectrum of the InAs/GaAs QD}

We performed calculations on a lens-shaped InAs/GaAs quantum dot
with 27.5 nm base diameter and 3.5 nm height.
The calculated transition energies of the exciton $X^0$, 
the charged excitons $X^-$ and $X^+$ and the
biexciton $XX$ of $S_h \rightarrow S_e$ transitions
are given in Fig.~\ref{fig:energy}.
The neutral $X^0$ has the largest transition energy of 981 meV, 
while charged excitons and biexcitons have 1 - 3 meV red shifts relative 
to the neutral exciton, due to correlation effects, \cite{bester03b}
in agreement with experiments. 
\cite{hartmann00,regelman01}
For the neutral exciton, the exciton line splits into four lines 
shown on the righthand side of Fig.~\ref{fig:energy},
as a consequence of the asymmetric 
electron-hole exchange energies. \cite{bester03}
The two low lying lines labeled $D1$, $D2$
are ``dark states'' (forbidden 
due to the angular momentum selection rule of the Bloch part of the 
electron/hole states), 
whereas the upper two states ($B1$, $B2$) are bright states. 
The calculated energy splitting between the bright and dark states 
is about 210 $\mu$eV.
The energy splitting between two dark states $D1$, $D2$ is about 1
$\mu$eV, whereas the
splitting between two bright states $B1$, $B2$ is about 10 $\mu$eV, 
in agreement
with previous calculations. \cite{bester03b}  
The low energy bright state $B1$ is polarized along the [110] direction, 
whereas the high energy state $B2$ is polarized along the 
[1$\bar{1}$0] direction.
Although the biexciton $XX$ states themselves have no fine structures, 
the $XX$ to $X$ transition exhibit fine structures. 
%which have revered order to those of $X$.  
The charged
excitons $X^+$ and $X^-$ have no fine structures.   

The energy of neutral exciton $P_h \rightarrow S_e$
transition is calculated here 995 meV, also shown
in Fig. \ref{fig:energy}. Like the  $S_h \rightarrow S_e$ transition, the
$P_h \rightarrow S_e$ transition also split into four fine structures. 
The two lower energy states are ``forbidden'' by the Bloch function
angular momentum selection rule, whereas
the two high energy states $B1$ and $B2$ are allowed
by the Bloch part.
In the far field measurement,
the $P_h \rightarrow S_e$ transition is forbidden for the lens-shaped QDs, 
due to the envelope function angular momentum selection rule (i.e.,
the overlap integrals
between $P_h$ and $S_e$ envelope functions are very small).
In practice, this transition is partially
allowed, \cite{narvaez06} but nevertheless very weak.
However, unlike the selection rules
for the Bloch functions,
the envelope function angular momentum selection rule does not apply
to NSOM, which involves only the local transition and does not
feel the global spatial symmetry,\cite{hohenester05} 
i.e., the overlap of $P_h$ and $S_e$ orbitals are not zero
at given point, and 
$P_h \rightarrow S_e$ transitions have comparable
intensity to the $S_h \rightarrow S_e$ transitions.
This will be further discussed in 
Sec.\ref{sec:p-s}.

\begin{figure}
\includegraphics[width=3.0in,angle=0]{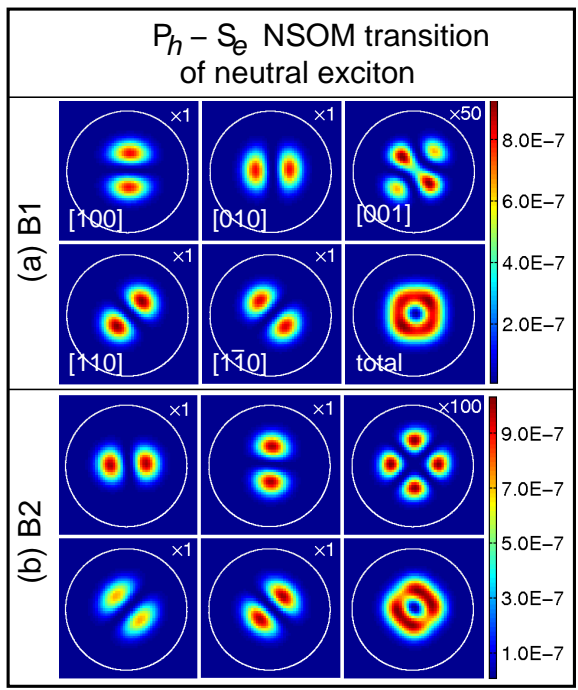}
\caption{
(Color online)
The NSOM of excitonic $P_h\rightarrow S_e$ transitions 
(see Fig. \ref{fig:energy}) in 
self-assembled InAs/GaAs QD with $b$=27.5 nm and $h$=3.5 nm, 
calculated from Eq. (\ref{eq:nsom-final}) for 
(a) $B1$ and (b) $B2$ states.% (see Fig. \ref{fig:energy}).
The unit of the color-bar is arbitrary but equal for all sub-figures.
In each sub-figure, there are six panels showing
the transition intensities of different polarizations. 
In the first row, the
polarizations are along the [100], [010] and [001] direction respectively,
and in the
second row, the polarizations are along the [110] and [1$\bar{1}$0]
directions. The last panels show 
the sum of transition intensities of different polarizations. 
We magnify the transition intensities 
by the factors on the upper-right corner of each
panel. 
The white circles (27.5 nm in diameter) show the boundaries the QDs.
}
\label{fig:1e1h-ps}
\end{figure}

\subsection{NSOM $S_h \rightarrow S_e$ transition for neutral exciton in
  the InAs/GaAs QD}
\label{sec:s-s}

The NSOM images of excitonic $S_h \rightarrow S_e$ transition of $X^0$
are shown in Fig.~\ref{fig:1e1h-ss}.
We show the images of all
fine structures components $B1$, $B2$, $D1$, $D2$
in Figs. \ref{fig:1e1h-ss}(a), 
\ref{fig:1e1h-ss}(b), 
\ref{fig:1e1h-ss}(c) and \ref{fig:1e1h-ss}(d)
respectively. 
The color-bars indicate the magnitudes of transition intensities.
We use arbitrary but equal units for the transition intensity
in all sub-figures.
Each sub-figure is composed of six panels showing the transition intensity of 
different polarizations.  
The first row in each sub-figure shows the transition 
of polarization along the [100],
[010] and [001] directions, whereas the second row of each sub-figure
shows the transition of
polarization along the [1$\bar{1}$0] and [110] directions. 
The final panel is the
total transition intensity, 
being the sum of intensities over polarization along the
[100], [010] and [001] directions.
(The sum of intensities over the [100], [010] polarization equals the sum over 
 the [1$\bar{1}$0] and [110] polarization.)
We magnified the transition intensity by the factor shown on 
the upper right corner of each panel,
so the transition intensities in all sub-figures can be in similar scale.

Considering first the total transition intensities,
we notice that the transitions originate 
from a relatively small region around the 
center of the dot.
The transition intensities of dark states $D1$ [Fig. \ref{fig:1e1h-ss}(a)]
and $D2$ [Fig. \ref{fig:1e1h-ss}(b)]
are about 2 - 3 orders of magnitude lower than the
those of bright exciton states $B1$ [Fig. \ref{fig:1e1h-ss}(c)]
and $B2$ [Fig. \ref{fig:1e1h-ss}(d)], because they
are ``forbidden'' due to angular momentum selection 
rule for the heavy-hole to conduction band transitions.
Therefore the shapes of $D1$ and $D2$ are determined by the 
shape of higher bands [see Eq. (\ref{eq:dipole-simple-delta})]. 
In contrast, the bright states $B1$ and $B2$ 
have $S$-like shape as expected.
However, the NSOM of $B1$ and $B2$ are not perfectly round, 
even though the lens-shaped QD we choose has 
cylindrical symmetry:  
we see that $B1$ is elongated along the [110] direction, whereas
$B2$ is elongated along the [0$\bar{1}$0] direction.

We next discuss the NSOM images of different polarizations,
focusing on the bright states.  
For far field excitation, the
low energy bright exciton $B1$ is calculated to be polarized along 
the [110] direction, whereas the high energy 
bright exciton $B2$ is polarized along the [1$\bar{1}$0] direction.
This is clearly seen in the NSOM plots of Fig. \ref{fig:1e1h-ss}, 
by comparing the [110] polarized
plots to the [1$\bar{1}$0] polarized plot.
However, NSOM can provide much more information than the usual far field
measurement:
The dipole moments have the following relations,
${\bf p}_{[110]}= ({\bf p}_{[100]} + {\bf p}_{[010]})/\sqrt{2} $ and
${\bf p}_{[1\bar{1}0]}= ({\bf p}_{[100]} - {\bf p}_{[010]})/\sqrt{2}$.
Therefore, by comparing the NSOM images of different polarization, we can
obtain the relative phases between ${\bf p}_{[100]}$ and ${\bf p}_{[010]}$
at each spatial point. 
For the bright states of 
$S_h \rightarrow S_e$ transitions, the relations are simple, i.e.,
${\bf p}_{[100]}$ and  ${\bf p}_{[010]}$ have almost 
fixed phases $\pi/2$ everywhere inside the QD.
As we know, the ${\bf p}_{[100]}$ corresponding to the $|X \rangle$ component 
in the hole Bloch functions, whereas ${\bf p}_{[010]}$ corresponding to 
the $|Y \rangle$ component. We therefore
obtain additional information about the
character of the hole Bloch functions as a function of position inside the QDs.

Figure \ref{fig:1e1h-ss} also 
show the NSOM image of polarization along the [001] direction. For the
bright exciton, the [001] components are
significantly weaker than the components polarized in 
(001)-plane, and may not be
detectable in practice. However, they are of theoretical interest. 
The polarization
along the [001] direction come from the emission of light hole. Therefore, the
[001] image shows how the light hole components distributed in the
QDs.

\begin{figure}
\includegraphics[width=3.0in,angle=0]{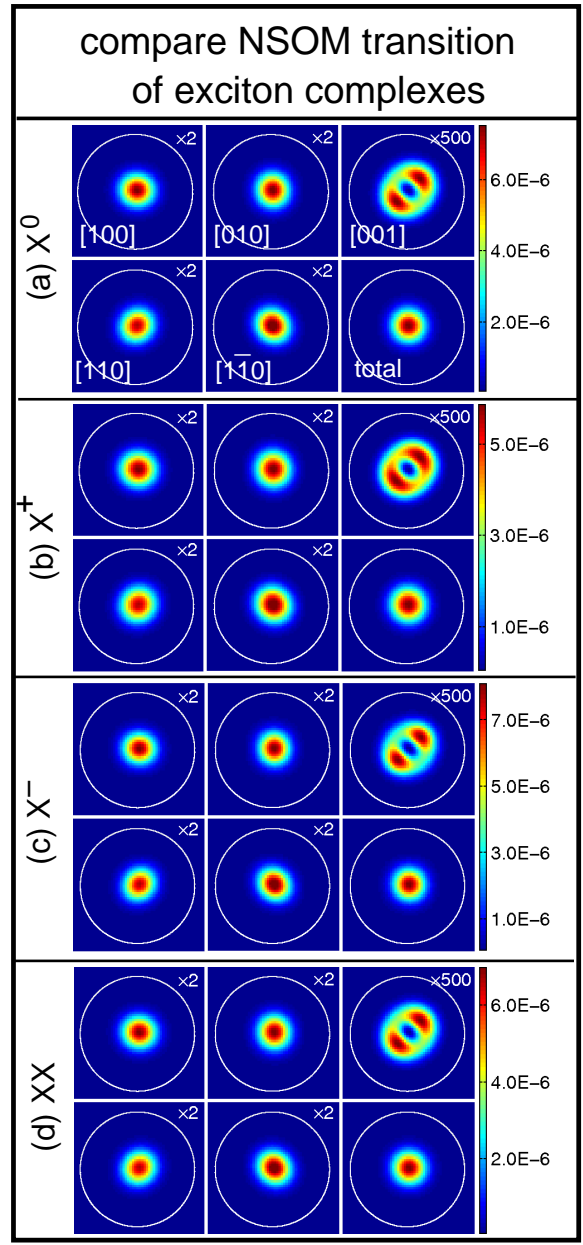}
\caption{
(Color online)
The total NSOM images (sum over all fine structures) 
of excitonic $S_h\rightarrow S_e$ transitions
(see Fig. \ref{fig:energy}) for 
(a) $X^0$, (b) $X^{+}$, (c) $X^-$, and (d) $XX$
in self-assembled InAs/GaAs QD with $b$=27.5 nm and $h$=3.5 nm, 
calculated from Eq. (\ref{eq:nsom-final}).
The unit of the color-bar is arbitrary but equal for all sub-figures.
In each sub-figure, there are six panels showing
the transition intensities of different polarizations. 
In the first row, the
polarizations are along the [100], [010] and [001] direction respectively,
and in the
second row, the polarizations are along the [110] and [1$\bar{1}$0]
directions. The last panels show %the total transition intensities being 
the sum of transition intensities of different polarizations.
We magnify the transition intensities 
by the factors on the upper-right corner of each
panel. 
The white circles (27.5 nm in diameter) show the boundaries the QDs.
}
\label{fig:compare}
\end{figure}
\subsection{NSOM $P_h \rightarrow S_e$ transition for neutral exciton in
  the InAs/GaAs QD}
\label{sec:p-s}      

Figure \ref{fig:1e1h-ps} depicts 
the $P_h \rightarrow S_e$ transitions, 
showing the two bright states ($B1$, $B2$) 
out of the four fine-structure split states. 
Similar to Fig. \ref{fig:1e1h-ss}, we show the NSOM images with 
polarization along the [100], 
[010], [001] directions
in the first row, and those with the [110], [1$\bar{1}$0] polarization 
and the total transition intensities in the second row.
We see that the $P_h \rightarrow S_e$ transitions have 
comparable intensities to those of the $S_h \rightarrow S_e$ transitions.
The shape of total transition intensity of the $B1$ state somewhat differs
from the intensity from the $B2$ state.
However, both states have ring like structures, 
and both states have two peaks in the [110] direction.

Each of the polarization-resolved plots looks like orbital-$p$ functions, 
although the
transition peaks are oriented in different directions.
For example, 
for the $B1$ state, the peaks of the [100] ([010]) polarized light
are oriented along the [100] ([010]) direction, 
whereas the peaks of [110] ([1$\bar{1}$0]) polarized light 
are oriented in [110] ([1$\bar{1}$0]) direction. 
As discussed in Sec.\ref{sec:s-s}, we can obtain the phase relations
of $|X \rangle$ and $|Y \rangle$ components in hole wavefunctions. 
We see that unlike the previous case involving an
$S_h$ wavefunction, in the Bloch part of the $P_h$
wavefunctions, $|X \rangle$ and $|Y \rangle$ components
have different phases at each
point of the QDs. Similar results are found for the $B2$ state.

\begin{figure}[h]
\includegraphics[width=2.5in,angle=-90]{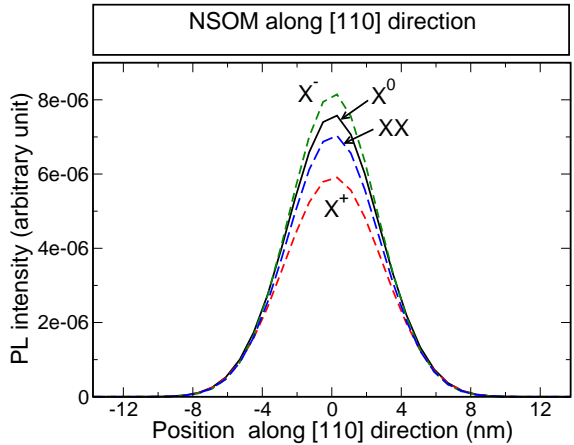}
\caption{ 
(Color online)
The NSOM transition intensities of $X^0$ 
(black solid line), $X^+$ (red dashed), $X^-$ (green dashed)
and $XX$ (blue dashed) along the [110]
direction in self-assembled InAs/GaAs QD 
with $b$=27.5 nm and $h$=3.5 nm, 
calculated from Eq. (\ref{eq:nsom-final}). 
The origin is at the center of the dot. 
} 
\label{fig:compare_line}
\end{figure}
\subsection{Comparison of NSOM of $X^0$, $X^+$, $X^-$ and $XX$ excitons in
  the InAs/GaAs QD}

In Figs. \ref{fig:compare} (b), (c), (d)  we depict the NSOM images of 
the charged exciton 
$X^+$, $X^-$ and the biexciton $XX$, respectively, and compare them to the 
$X^0$  $S_h \rightarrow S_e$ transition from Fig. \ref{fig:compare}(a).
We also plot the NSOM transition intensities along the [110] direction in 
Fig. \ref{fig:compare_line} for  $X^0$, $X^+$, $X^-$ and $XX$. 
The transition intensities of $X^0$, $X^+$, $X^-$ and $XX$ 
along [1$\bar{1}$0] direction are very similar to those along the
[110] direction, and therefore are not plotted.  
We see that $X^-$ has the strongest transition intensity, 
followed by the $X^0$, $XX$ and $X^+$.
Even though, the charged excitons and the biexciton have 1 - 3 meV red shifts 
relative to $X^0$, they have
very similar shapes and sizes to $X^0$. The only noticeable difference being that
$X^+$ has a slightly lager size than the other transitions.  
In Ref. \onlinecite{matsuda03}, it was found experimentally and confirmed
theoretically by single-band effective mass calculations
\cite{nair97,hohenester04} 
that the radius of 
the biexciton NSOM signal is smaller than the radius of $X^0$. 
In this calculation, 
we did not find a significant difference in size and shape 
between $X^0$ and $XX$. 
However, the QDs used in the experiments (created by composition 
fluctuations in a quantum well) are much larger 
($\sim$ 200 nm in diameter) than the self-assembled QDs we studied here,
therefore the 
correlation effects are stronger. 
On the other hand, we ignored the continuum states in the CI calculations, 
which may also underestimate the correlations effects to the wavefunctions.

\section{summary}

We calculate the NSOM images of exciton complexes 
in self-assembled InAs/GaAs QD using an atomistic pseudopotential method
followed by a configuration interaction treatment of the correlations. The
NSOM images reveal the spatial
structure of exciton and give informations about the Bloch function character
of the wavefunction, which remain unobserved in conventional far-field
measurements. We show that the $P_h \rightarrow S_e$ transition, which is
forbidden in the far field PL measurement, has comparable transition
intensity to that of the $S_h \rightarrow S_e$ transition in NSOM
measurements. We also calculate the NSOM image of charged exciton $X^+$,
$X^-$, and
biexciton $XX$, and compare them with the image of $X^0$. We found that
the images obtained for charged excitions and the biexcition show very similar
features to the ones obtained from the exciton decay.

%%%%%%%%%%%%%%%%%%%%%%%%%%%%%%%%%%%%%%%%
\acknowledgments
%%%%%%%%%%%%%%%%%%%%%%%%%%%%%%%%%%%%%%%%

This work is supported by the Chinese National
Fundamental Research Program, the Innovation
funds and ``Hundreds of Talents'' program from Chinese Academy of
Sciences, and National Natural Science Foundation of China (Grant
No. 10674124).
The work done at NREL was funded by the U.S. Department of Energy, 
Office of Science,
Basic Energy Science, Materials Sciences and Engineering, LAB-17 initiative,
under Contract No. DE-AC36-99GO10337 to NREL.


\begin{thebibliography}{22}
\expandafter\ifx\csname natexlab\endcsname\relax\def\natexlab#1{#1}\fi
\expandafter\ifx\csname bibnamefont\endcsname\relax
  \def\bibnamefont#1{#1}\fi
\expandafter\ifx\csname bibfnamefont\endcsname\relax
  \def\bibfnamefont#1{#1}\fi
\expandafter\ifx\csname citenamefont\endcsname\relax
  \def\citenamefont#1{#1}\fi
\expandafter\ifx\csname url\endcsname\relax
  \def\url#1{\texttt{#1}}\fi
\expandafter\ifx\csname urlprefix\endcsname\relax\def\urlprefix{URL }\fi
\providecommand{\bibinfo}[2]{#2}
\providecommand{\eprint}[2][]{\url{#2}}

\bibitem[{\citenamefont{Michler et~al.}(2000)\citenamefont{Michler, Kiraz,
  Becher, Schoenfeld, Petroff, Zhang, Hu, and Imamo\u{g}lu}}]{michler00}
\bibinfo{author}{\bibfnamefont{P.}~\bibnamefont{Michler}},
  \bibinfo{author}{\bibfnamefont{A.}~\bibnamefont{Kiraz}},
  \bibinfo{author}{\bibfnamefont{C.}~\bibnamefont{Becher}},
  \bibinfo{author}{\bibfnamefont{W.~V.} \bibnamefont{Schoenfeld}},
  \bibinfo{author}{\bibfnamefont{P.~M.} \bibnamefont{Petroff}},
  \bibinfo{author}{\bibfnamefont{L.}~\bibnamefont{Zhang}},
  \bibinfo{author}{\bibfnamefont{E.}~\bibnamefont{Hu}}, \bibnamefont{and}
  \bibinfo{author}{\bibfnamefont{A.}~\bibnamefont{Imamo\u{g}lu}},
  \bibinfo{journal}{Science} \textbf{\bibinfo{volume}{290}},
  \bibinfo{pages}{2282} (\bibinfo{year}{2000}).

\bibitem[{\citenamefont{Akopian et~al.}(2006)\citenamefont{Akopian, Lindner,
  Poem, Berlatzky, Avron, Gershoni, Gerardot, and Petroff}}]{akopian06}
\bibinfo{author}{\bibfnamefont{N.}~\bibnamefont{Akopian}},
  \bibinfo{author}{\bibfnamefont{N.~H.} \bibnamefont{Lindner}},
  \bibinfo{author}{\bibfnamefont{E.}~\bibnamefont{Poem}},
  \bibinfo{author}{\bibfnamefont{Y.}~\bibnamefont{Berlatzky}},
  \bibinfo{author}{\bibfnamefont{J.}~\bibnamefont{Avron}},
  \bibinfo{author}{\bibfnamefont{D.}~\bibnamefont{Gershoni}},
  \bibinfo{author}{\bibfnamefont{B.~D.} \bibnamefont{Gerardot}},
  \bibnamefont{and} \bibinfo{author}{\bibfnamefont{P.~M.}
  \bibnamefont{Petroff}}, \bibinfo{journal}{Phys. \ Rev. \ Lett.}
  \textbf{\bibinfo{volume}{96}}, \bibinfo{pages}{130501}
  (\bibinfo{year}{2006}).

\bibitem[{\citenamefont{Gammon et~al.}(1996)\citenamefont{Gammon, Snow,
  Shanabrook, Katzer, and Park}}]{gammon96a}
\bibinfo{author}{\bibfnamefont{D.}~\bibnamefont{Gammon}},
  \bibinfo{author}{\bibfnamefont{E.~S.} \bibnamefont{Snow}},
  \bibinfo{author}{\bibfnamefont{B.~V.} \bibnamefont{Shanabrook}},
  \bibinfo{author}{\bibfnamefont{D.~S.} \bibnamefont{Katzer}},
  \bibnamefont{and} \bibinfo{author}{\bibfnamefont{D.}~\bibnamefont{Park}},
  \bibinfo{journal}{Phys.\ Rev.\ Lett.} \textbf{\bibinfo{volume}{76}},
  \bibinfo{pages}{3005} (\bibinfo{year}{1996}).

\bibitem[{\citenamefont{Bayer et~al.}(1999)\citenamefont{Bayer, Kuther,
  Forchel, Gorbunov, Timofeev, Sch\"afer, Reithmaier, Reinecke, and
  Walck}}]{bayer99}
\bibinfo{author}{\bibfnamefont{M.}~\bibnamefont{Bayer}},
  \bibinfo{author}{\bibfnamefont{A.}~\bibnamefont{Kuther}},
  \bibinfo{author}{\bibfnamefont{A.}~\bibnamefont{Forchel}},
  \bibinfo{author}{\bibfnamefont{A.}~\bibnamefont{Gorbunov}},
  \bibinfo{author}{\bibfnamefont{V.~B.} \bibnamefont{Timofeev}},
  \bibinfo{author}{\bibfnamefont{F.}~\bibnamefont{Sch\"afer}},
  \bibinfo{author}{\bibfnamefont{J.~P.} \bibnamefont{Reithmaier}},
  \bibinfo{author}{\bibfnamefont{T.~L.} \bibnamefont{Reinecke}},
  \bibnamefont{and} \bibinfo{author}{\bibfnamefont{S.~N.} \bibnamefont{Walck}},
  \bibinfo{journal}{Phys. Rev. Lett.} \textbf{\bibinfo{volume}{82}},
  \bibinfo{pages}{1748} (\bibinfo{year}{1999}).

\bibitem[{\citenamefont{Urbaszek et~al.}(2003)\citenamefont{Urbaszek,
  Warburton, Karrai, Gerardot, Petroff, and Garcia}}]{urbaszek03}
\bibinfo{author}{\bibfnamefont{B.}~\bibnamefont{Urbaszek}},
  \bibinfo{author}{\bibfnamefont{R.~J.} \bibnamefont{Warburton}},
  \bibinfo{author}{\bibfnamefont{K.}~\bibnamefont{Karrai}},
  \bibinfo{author}{\bibfnamefont{B.~D.} \bibnamefont{Gerardot}},
  \bibinfo{author}{\bibfnamefont{P.~M.} \bibnamefont{Petroff}},
  \bibnamefont{and} \bibinfo{author}{\bibfnamefont{J.~M.}
  \bibnamefont{Garcia}}, \bibinfo{journal}{Phys. Rev. Lett.}
  \textbf{\bibinfo{volume}{90}}, \bibinfo{pages}{247403}
  (\bibinfo{year}{2003}).

\bibitem[{\citenamefont{Regelman et~al.}(2001)\citenamefont{Regelman, Dekel,
  Gershoni, Ehrenfreund, Williamson, Shumway, Zunger, Schoenfeld, and
  Petroff}}]{regelman01}
\bibinfo{author}{\bibfnamefont{D.~V.} \bibnamefont{Regelman}},
  \bibinfo{author}{\bibfnamefont{E.}~\bibnamefont{Dekel}},
  \bibinfo{author}{\bibfnamefont{D.}~\bibnamefont{Gershoni}},
  \bibinfo{author}{\bibfnamefont{E.}~\bibnamefont{Ehrenfreund}},
  \bibinfo{author}{\bibfnamefont{A.~J.} \bibnamefont{Williamson}},
  \bibinfo{author}{\bibfnamefont{J.}~\bibnamefont{Shumway}},
  \bibinfo{author}{\bibfnamefont{A.}~\bibnamefont{Zunger}},
  \bibinfo{author}{\bibfnamefont{W.~V.} \bibnamefont{Schoenfeld}},
  \bibnamefont{and} \bibinfo{author}{\bibfnamefont{P.~M.}
  \bibnamefont{Petroff}}, \bibinfo{journal}{Phys.\ Rev.\ B}
  \textbf{\bibinfo{volume}{64}}, \bibinfo{pages}{165301}
  (\bibinfo{year}{2001}).

\bibitem[{\citenamefont{Hartmann et~al.}(2000)\citenamefont{Hartmann, Ducommun,
  Kapon, Hohenester, and Molinari}}]{hartmann00}
\bibinfo{author}{\bibfnamefont{A.}~\bibnamefont{Hartmann}},
  \bibinfo{author}{\bibfnamefont{Y.}~\bibnamefont{Ducommun}},
  \bibinfo{author}{\bibfnamefont{E.}~\bibnamefont{Kapon}},
  \bibinfo{author}{\bibfnamefont{U.}~\bibnamefont{Hohenester}},
  \bibnamefont{and} \bibinfo{author}{\bibfnamefont{E.}~\bibnamefont{Molinari}},
  \bibinfo{journal}{Phys.\ Rev.\ Lett.} \textbf{\bibinfo{volume}{84}},
  \bibinfo{pages}{5648} (\bibinfo{year}{2000}).

\bibitem[{\citenamefont{Bryant}(1998)}]{bryant98}
\bibinfo{author}{\bibfnamefont{G.~W.} \bibnamefont{Bryant}},
  \bibinfo{journal}{Appl. \ Phys. \ Lett.} \textbf{\bibinfo{volume}{72}},
  \bibinfo{pages}{768} (\bibinfo{year}{1998}).

\bibitem[{\citenamefont{Matsuda et~al.}(2003)\citenamefont{Matsuda, Saiki,
  Nomura, Mihara, Aoyagi, Nair, and Takagahara}}]{matsuda03}
\bibinfo{author}{\bibfnamefont{K.}~\bibnamefont{Matsuda}},
  \bibinfo{author}{\bibfnamefont{T.}~\bibnamefont{Saiki}},
  \bibinfo{author}{\bibfnamefont{S.}~\bibnamefont{Nomura}},
  \bibinfo{author}{\bibfnamefont{M.}~\bibnamefont{Mihara}},
  \bibinfo{author}{\bibfnamefont{Y.}~\bibnamefont{Aoyagi}},
  \bibinfo{author}{\bibfnamefont{S.}~\bibnamefont{Nair}}, \bibnamefont{and}
  \bibinfo{author}{\bibfnamefont{T.}~\bibnamefont{Takagahara}},
  \bibinfo{journal}{Phys. \ Rev. \ Lett.} \textbf{\bibinfo{volume}{91}},
  \bibinfo{pages}{177401} (\bibinfo{year}{2003}).

\bibitem[{\citenamefont{Flack et~al.}(1996)\citenamefont{Flack, Samarth,
  Nikitin, Crowell, Shi, Levy, and Awschalom}}]{flack96}
\bibinfo{author}{\bibfnamefont{F.}~\bibnamefont{Flack}},
  \bibinfo{author}{\bibfnamefont{N.}~\bibnamefont{Samarth}},
  \bibinfo{author}{\bibfnamefont{V.}~\bibnamefont{Nikitin}},
  \bibinfo{author}{\bibfnamefont{P.~A.} \bibnamefont{Crowell}},
  \bibinfo{author}{\bibfnamefont{J.}~\bibnamefont{Shi}},
  \bibinfo{author}{\bibfnamefont{J.}~\bibnamefont{Levy}}, \bibnamefont{and}
  \bibinfo{author}{\bibfnamefont{D.~D.} \bibnamefont{Awschalom}},
  \bibinfo{journal}{Phys. Rev. B} \textbf{\bibinfo{volume}{54}},
  \bibinfo{pages}{R17312} (\bibinfo{year}{1996}).

\bibitem[{\citenamefont{Hohenester et~al.}(2005)\citenamefont{Hohenester,
  Goldoni, and Molinari}}]{hohenester05}
\bibinfo{author}{\bibfnamefont{U.}~\bibnamefont{Hohenester}},
  \bibinfo{author}{\bibfnamefont{G.}~\bibnamefont{Goldoni}}, \bibnamefont{and}
  \bibinfo{author}{\bibfnamefont{E.}~\bibnamefont{Molinari}},
  \bibinfo{journal}{Phys. \ Rev. \ Lett.} \textbf{\bibinfo{volume}{95}},
  \bibinfo{pages}{216802} (\bibinfo{year}{2005}).

\bibitem[{\citenamefont{Williamson et~al.}(2000)\citenamefont{Williamson, Wang,
  and Zunger}}]{williamson00}
\bibinfo{author}{\bibfnamefont{A.~J.} \bibnamefont{Williamson}},
  \bibinfo{author}{\bibfnamefont{L.-W.} \bibnamefont{Wang}}, \bibnamefont{and}
  \bibinfo{author}{\bibfnamefont{A.}~\bibnamefont{Zunger}},
  \bibinfo{journal}{Phys.\ Rev.\ B} \textbf{\bibinfo{volume}{62}},
  \bibinfo{pages}{12963} (\bibinfo{year}{2000}).

\bibitem[{\citenamefont{Franceschetti et~al.}(1999)\citenamefont{Franceschetti,
  Fu, Wang, and Zunger}}]{franceschetti99}
\bibinfo{author}{\bibfnamefont{A.}~\bibnamefont{Franceschetti}},
  \bibinfo{author}{\bibfnamefont{H.}~\bibnamefont{Fu}},
  \bibinfo{author}{\bibfnamefont{L.-W.} \bibnamefont{Wang}}, \bibnamefont{and}
  \bibinfo{author}{\bibfnamefont{A.}~\bibnamefont{Zunger}},
  \bibinfo{journal}{Phys.\ Rev.\ B} \textbf{\bibinfo{volume}{60}},
  \bibinfo{pages}{1819} (\bibinfo{year}{1999}).

\bibitem[{\citenamefont{Keating}(1966)}]{keating66}
\bibinfo{author}{\bibfnamefont{P.~N.} \bibnamefont{Keating}},
  \bibinfo{journal}{Phys. Rev.} \textbf{\bibinfo{volume}{145}},
  \bibinfo{pages}{637} (\bibinfo{year}{1966}).

\bibitem[{\citenamefont{Martins and Zunger}(1984)}]{martins84}
\bibinfo{author}{\bibfnamefont{J.~L.} \bibnamefont{Martins}} \bibnamefont{and}
  \bibinfo{author}{\bibfnamefont{A.}~\bibnamefont{Zunger}},
  \bibinfo{journal}{Phys.\ Rev.\ B} \textbf{\bibinfo{volume}{30}},
  \bibinfo{pages}{R6217} (\bibinfo{year}{1984}).

\bibitem[{\citenamefont{Wang and Zunger}(1999)}]{wang99b}
\bibinfo{author}{\bibfnamefont{L.-W.} \bibnamefont{Wang}} \bibnamefont{and}
  \bibinfo{author}{\bibfnamefont{A.}~\bibnamefont{Zunger}},
  \bibinfo{journal}{Phys.\ Rev.\ B} \textbf{\bibinfo{volume}{59}},
  \bibinfo{pages}{15806} (\bibinfo{year}{1999}).

\bibitem[{\citenamefont{Dragoman and Dragoman}(2002)}]{Dragoman_book}
\bibinfo{author}{\bibfnamefont{D.}~\bibnamefont{Dragoman}} \bibnamefont{and}
  \bibinfo{author}{\bibfnamefont{M.}~\bibnamefont{Dragoman}},
  \emph{\bibinfo{title}{Optical characterization of solids}}
  (\bibinfo{publisher}{Springer Verlag}, \bibinfo{year}{2002}).

\bibitem[{\citenamefont{Nair and Takagahara}(1997)}]{nair97}
\bibinfo{author}{\bibfnamefont{S.~V.} \bibnamefont{Nair}} \bibnamefont{and}
  \bibinfo{author}{\bibfnamefont{T.}~\bibnamefont{Takagahara}},
  \bibinfo{journal}{Phys. Rev. B} \textbf{\bibinfo{volume}{55}},
  \bibinfo{pages}{5153} (\bibinfo{year}{1997}).

\bibitem[{\citenamefont{Hohenester et~al.}(2004)\citenamefont{Hohenester,
  Goldoni, and Molinari}}]{hohenester04}
\bibinfo{author}{\bibfnamefont{U.}~\bibnamefont{Hohenester}},
  \bibinfo{author}{\bibfnamefont{G.}~\bibnamefont{Goldoni}}, \bibnamefont{and}
  \bibinfo{author}{\bibfnamefont{E.}~\bibnamefont{Molinari}},
  \bibinfo{journal}{Appl. \ Phys. \ Lett.} \textbf{\bibinfo{volume}{84}},
  \bibinfo{pages}{3963} (\bibinfo{year}{2004}).

\bibitem[{\citenamefont{Bester and Zunger}(2003)}]{bester03b}
\bibinfo{author}{\bibfnamefont{G.}~\bibnamefont{Bester}} \bibnamefont{and}
  \bibinfo{author}{\bibfnamefont{A.}~\bibnamefont{Zunger}},
  \bibinfo{journal}{Phys. Rev. B} \textbf{\bibinfo{volume}{68}},
  \bibinfo{pages}{073309} (\bibinfo{year}{2003}).

\bibitem[{\citenamefont{Bester et~al.}(2003)\citenamefont{Bester, Nair, and
  Zunger}}]{bester03}
\bibinfo{author}{\bibfnamefont{G.}~\bibnamefont{Bester}},
  \bibinfo{author}{\bibfnamefont{S.~V.} \bibnamefont{Nair}}, \bibnamefont{and}
  \bibinfo{author}{\bibfnamefont{A.}~\bibnamefont{Zunger}},
  \bibinfo{journal}{Phys. Rev. B} \textbf{\bibinfo{volume}{67}},
  \bibinfo{pages}{161306} (\bibinfo{year}{2003}).

\bibitem[{\citenamefont{Narvaez and Zunger}(2006)}]{narvaez06}
\bibinfo{author}{\bibfnamefont{G.~A.} \bibnamefont{Narvaez}} \bibnamefont{and}
  \bibinfo{author}{\bibfnamefont{A.}~\bibnamefont{Zunger}},
  \bibinfo{journal}{Phys. \ Rev. \ B} \textbf{\bibinfo{volume}{74}},
  \bibinfo{pages}{045316} (\bibinfo{year}{2006}).

\end{thebibliography}
 \end{document}